\documentclass[runningheads,a4paper,oribibl]{llncs}

\usepackage{amssymb}
\usepackage{graphicx}
\usepackage{wrapfig}

\usepackage{url}
\urldef{\mailsa}\path|{marcos.baezgonzalez, chiara.dalpiaz, fatbardha.hoxha, alessia.tovo,|  
\urldef{\mailsb}\path|valentina.caforio, fabio.casati}@unitn.it|    
\newcommand{\keywords}[1]{\par\addvspace\baselineskip
\noindent\keywordname\enspace\ignorespaces#1}

\newcommand\blfootnote[1]{%
  \begingroup
  \renewcommand\thefootnote{}\footnote{#1}%
  \addtocounter{footnote}{-1}%
  \endgroup
}

\begin{document}

\mainmatter  

\title{Personalized Persuasion for Social Interactions in Nursing Homes}

\titlerunning{Personalized Persuasion for Social Interactions in Nursing Homes}

%
%
\author{Marcos Baez%
\and Chiara Dalpiaz\and Fatbardha Hoxha\and Alessia Tovo\and \\
Valentina Caforio \and Fabio Casati}
\authorrunning{Personalized Persuasion for Social Interactions in Nursing Homes}

\institute{Dipartimento di Ingegneria e Scienza dell’Informazione,\\
University of Trento, Italy\\
\mailsa\\
\mailsb\\
}

%
%

\toctitle{Lecture Notes in Computer Science}
\tocauthor{Authors' Instructions}
\maketitle

\begin{abstract}
This paper presents our preliminary investigation and approach towards a mixed physical-virtual technology for stimulating social interactions among and with older adults in nursing homes. We report on set of surveys, apps and focus groups aiming at understanding the different motivations and obstacles in promoting social interactions in institutionalised care. We then present our approach to address some of the key themes found, e.g., the technological disparity, lack of conversation topics and opportunities to interact.\blfootnote{
\begin{wrapfigure}{l}{0.12\textwidth}
  \vspace{-44pt}
  \begin{center}
    \includegraphics[height=1cm]{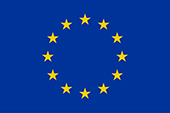}
  \end{center}
    \vspace{-30pt}
\end{wrapfigure}
{\fontfamily{cmss}\selectfont
\noindent This project has received funding from the European Union's Horizon 2020 research and innovation programme under the Marie Skłodowska-Curie grant agreement No 690962. \\
}
}

\keywords{social interactions, older adults, nursing homes, persuasion strategies}
\end{abstract}

\section{Introduction}

The transition to residential care is one of the most difficult experiences in the life of elderly and the family, requiring the adaptation to a completely new personal and social context \cite{lee2002review}.  Family involvement, and especially maintaining an emotional bond through visits and family updates, is important to preserve the resident’s sense of stability and connectedness along this process \cite{friedemann1997family}.  Social integration with peers is also largely regarded as beneficial. Connecting to others helps in the adaptation, can foster friendships and sense of belonging, and has been found to be one of the key elements contributing to the quality of life in residential care \cite{bradshaw2012living}. In contrast, failing to keep socially active contributes to feelings of loneliness, boredom, and helplessness, regarded as the plagues of nursing home life \cite{thomas1996life,lindley2015placing}.

Different physical, psychological and contextual factors influence the opportunities and motivations of older adults to interact with others. Poor physical health and mental illness can pose a barrier to participating in social activities thus contributing to social isolation \cite{edelbrock2001social}. People with poor psychological resources, socially intimidated, or with fear of intimacy might also find difficulties developing relationships and feeling connected \cite{windle2004variations,rokach1996causes}. Other factors such as kinship and social network closeness have also been found to be related to level of visits and calls by family \cite{port2001resident}. All these factors explain the effectiveness of interventions aiming at reducing social isolation targeting specific groups \cite{cattan2005preventing}.

Technology supporting social interactions in this context has largely focused on \emph{enabling} social interactions, exploring technologies ranging from email and video calls to social networks \cite{findlay2003interventions}, without considering the personal traits and the context that determine the user intentions, abilities and motivations to interact. 

In this research project we aim at addressing the above scenario by providing the technology that can \emph{enable} and \emph{motivate} social interaction of the nursing home residents with peers and family. We recognize the potential of persuasion technologies \cite{oinas2009persuasive} as an instrument to empower social interactions by creating sustainable and more meaningful conversations. However, for the same reasons explained before, we believe that the vision that \emph{one-design-fits-all} would not be feasible \cite{orji2014design}, and thus, we set to explore the tailoring of tools and persuasion strategies to the intentions, motivations, preferences and needs of the various actors.

In what follows we report on the early steps towards this idea, with the design of a tailored newspaper, where the content is generated semi-automatically with contributions from older adults and family members, and that serves as a tool for creating meaningful conversations and ultimately increasing the wellbeing of the older adults and their interactions with children and grandchildren. The design incorporates ideas from the reminiscence therapy and crowdsourcing, with tasks that are sensitive to the motivations of the various actors.  The project is in collaboration with the leading technology provider for nursing homes in Italy and with the nursing homes of the Trentino province in Italy.

\section{Problem exploration}

To gain a deeper understanding of the problem we developed and run a set of surveys, apps and focus groups to get qualitative and quantitative information on intergenerational interactions, with particular attention (for the focus groups in particular) to interactions in institutionalized care.
A first survey aimed at understanding interactions (and, specifically, the reasons for non-interaction) among young adults and their grandparents. We collected 100 responses from a convenience sample of university students in Trento, Italy. The results show that 
\begin{enumerate}
\item The majority of young adults have rather infrequent contact with some grandparents (82\% report physical and phone contacts of less than once a week, 52\% reports no contact at all in the last month)
\item The reasons for infrequent phone contact are lack of time (55\%) and lack of common topics of conversation (also 55\%). 23\% of respondents also cite cognitive difficulties as a limiting factor as well as pain in thinking at the relative in a physically or cognitively challenged condition.
\end{enumerate}

A second survey focused on those cases where interactions with grandparents was more frequent, to understand which are the activities they do together (in case of F2F meetings) or which are the common topics of conversation. A similar convenience sample also of 100 respondents, suggested that the most common activities are eating and watching TV (69 and 43\% respectively) followed by reading about recent news, home activities (cleaning) and watching pictures. Common topics of conversations includes progress in school (70\%), family news (68\%), health condition of the grandparent (55\%), and food (41\%).	

An additional broader set of surveys asked young adults to look at their Facebook posts and state for each of them if they would share them with their grandparents if it was easy to do so, or if not, why. We collected results for over 2000 posts with respondents from around the world, and the result we got quite consistently is that younger adults would feel comfortable in sharing their picture-based posts (over 75\% of them) with grandparents. Only for less than 25\% of the picture-based posts do people feel that they prefer not to share for privacy reasons, that grandparents would not understand them, or that would not be interested in them. The intention to share went significantly down (to less than 50\%) for posts related to link or status updates.

Given that interactions with grandchildren is a major source of joy for older adults, this picture points to the need for stimulating interactions, and it suggests that the key to doing so lies in i) creating common topics of conversation (with specific ideas on which ones these may be, such as family news, photos, or food) and ii) trying to initiate a “conversation” by making it, at least at first, almost quick and effortless for young adults to interact.

Focus groups conducted in three nursing homes in the Trentino province with staff and relatives helped understand more about the interactions among nursing home residents as well as between residents and their children, who are often their primary caretakers. What we learned is that:
\begin{enumerate}
\item Friendship relations are difficult in a nursing homes: people are not there by choice and do not see things in common with the other residents, besides the fact that they all need assistance and are in the same nursing home.
\item The primary caretaker spends a lot of time in the nursing home and visit very frequently – often they do so daily. However, many members of the immediate family (brothers or other children of the resident) are more absent and this causes frustration and even anger in both the resident and the primary caretaker.
\item Grandchildren rarely visit. This does not cause anger in the primary caretaker, but the caretaker tries to involve them and make them interested often failing to do so.
\end{enumerate}

The take-home message from the focus group is that an opportunity to improve the social interactions and quality of relations for people in nursing homes and their families is to persuade other family members to visit. In addition, the large amount of time spent by the primary caretaker in the nursing homes means that we have the opportunity to \emph{leverage} their presence there to facilitate IT-mediated interaction with other family members even in the (quite common) case that the resident lacks the cognitive ability to carry out the virtual interaction.

\section{Users, groups, and objectives}
From our survey and analysis of the literature we identified groups of users, the behaviors that we need to motivate and the potential barriers that they face. These aspects are summarized in Table \ref{table:groups}.

For each group, different characteristics define their ability and willingness to interact, which we take as the starting point:
\begin{itemize}  
\item For residents in the nursing homes we are collaborating with, distinction is made based on the type of care, defining subgroups such as intensive care, Alzheimer and (semi)independent; 
\item Children behavior seems to be related to being the main caretaker in the family, and
\item Grandchildren, behavior seems to be related to previous living arrangements, i.e., whether it was living with the grandparents, and distance.
\end{itemize}

\begin{table}[t!]
\small
\caption{Messages coding scheme}
\label{table:groups}  
\begin{tabular}{ |p{0.2\columnwidth}| p{0.35\columnwidth} | p{0.45\columnwidth} |} 
\hline
 & Target behavior & Constraints \\ 
\hline

Residents &
Physical interaction with peers &
Physical, mental or psychological factors, lack of opportunities, lack of conversation topics, perceived lack of common history and topics of conversations  \\
\hline

Children &
Virtual interactions with parents
Physical interactions with parents (visits) &
Distance, lack of time, suffering resulting from seeing parents in difficult condition, disinformation, delegation of responsibilities \\
\hline

Grandchildren &
Virtual interactions with the grandparents &
Lack of time, lack of conversation topics, lack of awareness of the positive impact \\

\hline
\end{tabular}  
\end{table}

All the above actors are clearly defined by particular interests (topics), which also define the level of affinity. 

\section{Approach and work in progress}
In this project we enable and stimulate interaction by combining: 
\begin{enumerate}
\item	Technologies for semi-automatically building a physical newspaper that includes “family news” by enhancing and curating information available from social networks
\item	Reminiscence techniques that, in addition to the intrinsic benefits of reminiscence, give us a way to collect information helpful to find common stories among residents
\item	Middleware technologies to simplify and mediate interaction among ad hoc applications to be used in nursing homes with social network commonly used by children and most of all by grand children. 
\end{enumerate}

\begin{figure}[t!]
\centering
\includegraphics[width=\columnwidth]{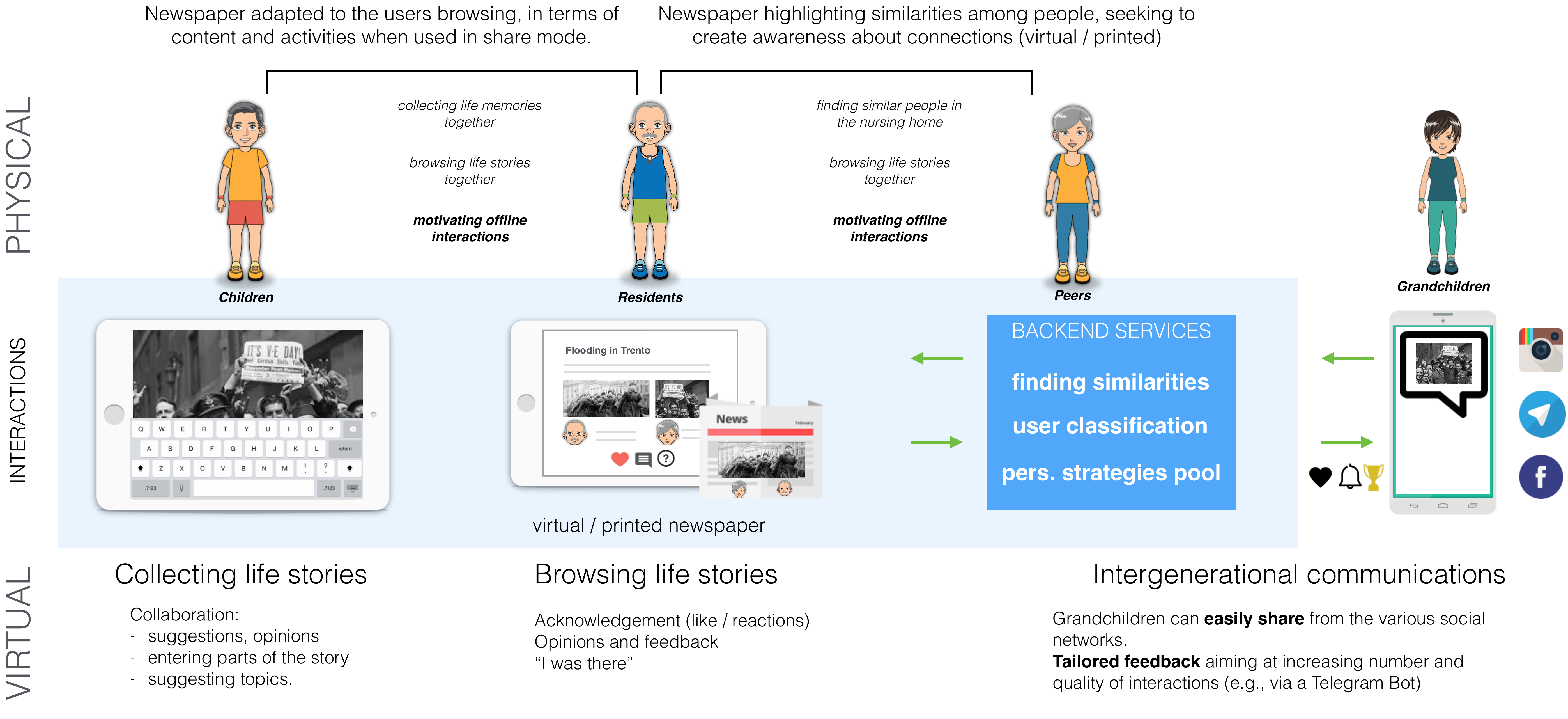}
\caption{Technological scenario of the approach}
\label{fig:architecture}
\end{figure}

The basic idea and usage patterns consist of collecting information (pictures, life stories, etc) from older adults with the help of the visiting children (the primary caretaker) or volunteers, through an application designed for this purpose. This information is processed to find common aspects with the life of other residents (people who lived the same experiences, or in the same city), thereby providing the opportunity for creating a bonding and initiating therefore a conversation. Information semi-automatically obtained from social networks (typically from Instagram accounts) of children and grandchildren can also be used to derive common topics of conversation (through common aspects of the life of the children or grandchildren). The usage scenario is illustrated in Figure \ref{fig:architecture}.

The newspaper is the primary interface towards the resident and is provided in both physical and virtual form. It contains snippets of the life of relatives of residents as well as reports on life in nursing homes and, weekly, feature stories about one resident.
All news items are related to one or more residents and, where possible, point to possible commonalities among residents. 
The newspaper is also meant as an instrument that makes visit more fun and as a way to bond children (as commonalities are also through children) so that they also are enticed to visit more often – especially children besides the primary caretaker. 

Finally, collecting pictures about life stories, and also enabling residents to view and talk about what they see about grandchildren in the newspaper is also an opportunity to involve grandchildren in a conversation, by probing them with snippets of information or feedback through a channel they already use (chat software or Instagram/Facebook), without the need of a new app on their side that they would need to install and open.

In general, what we are aiming at is the creation of topics of conversations, the creation of awareness of common history and common aspects in life, and the reduction of barriers to interaction. These aspects play out differently for the three target groups, with common history being instrumental to persuasion to “make friends” among residents and to make visits more fun for children, and reduction of barriers more oriented towards resident-grandchildren interaction.  

We believe this technological support provides the environment that will enable the different actors to develop stronger and more meaningful conversations. The personalization we are aiming at, however, is not limited to content and activities, but extends to the strategies that will enable sustainable contributions and interactions. Our hypothesis based on our preliminary work that includes a literature review, surveys and focus groups, is that the support will range from simple enabler to coaching mechanisms sensitive to the characteristics of the peers in a conversation.  Further studies will help us identify the subgroups of users, select the persuasion strategies to use in the personalisation, and understand what strategies work and for whom. 
\\

\noindent \textbf{Acknowledgment}. This project has received funding from the European Union's Horizon 2020 research and innovation programme under the Marie Skłodowska-Curie grant agreement No 690962. This work was also supported by the project ``Evaluation and enhancement of social, economic and emotional wellbeing of older adults" under the agreement no. 14.Z50.310029, Tomsk Polytechnic University" and by the Trentino project ``Collegamenti".

\bibliographystyle{splncs03}
\bibliography{references}

\end{document}